# Cross-layer Design in Cognitive Radio Standards

Vahid Towhidlou, Mohammad Shikh-Bahaei

Centre for Telecommunications Research

King's College London, UK

ABSTRACT

**The growing demand for wireless applications and services on the one hand, and limited available radio spectrum on the other hand have made cognitive radio (CR) a promising solution for future mobile networks. It has attracted considerable attention by academia and industry since its introduction in 1999 and several relevant standards have been developed within the last decade. Cognitive radio is based on four main functions, spanning across more than one layer of OSI model. Therefore, solutions based on cognitive radio technology require cross layer (CL) designs for optimum performance. This article briefly reviews the basics of cognitive radio technology as an introduction and highlights the need for cross layer design in systems deploying CR technology. Then some of the published standards with CL characteristics are outlined in a later section, and in the final section some research examples of cross layer design ideas based on the existing CR standards conclude this article.**

## I. INTRODUCTION

Over the past decade mobile services have evolved from basic voice communication to mobile broadband multimedia services and sophisticated wireless applications. Today, we cannot imagine a world without wireless communications and smart mobile devices. Any wireless communication relies on the availability of radiofrequency spectrum, which is inherently a finite resource and cannot be produced. This unprecedented growth of wireless devices and services has motivated academia and industry to look for new solutions in utilizing the wireless spectrum more intelligently and efficiently.

Studies of the spectrum scarcity problem by various regulatory bodies around the globe, including the federal communication commission (FCC) in the United States of America and OfCom in the United Kingdom have indicated that the licensed spectrum bands are severely under-utilized at any given time and location [1] and the overall spectrum utilization is less than 15%, mainly due to the traditional static spectrum allocation regulation [1].

The concept of cognitive radio (CR), first proposed in 1999 by Joseph Mitola III [2], opened up a new way of better utilization of the scarce wireless spectrum resource. In fact, the original idea introduced by Mitola embraced a broader perspective, and efficient utilization of wireless spectrum is one of the applications of a cognitive system;

> *"The point in which wireless personal digital assistants (PDAs) and the related networks are sufficiently computationally intelligent about radio resources and related computer-to-computer communications to detect user communications needs as a function of use context, and to provide radio resources and wireless services most appropriate to those needs"* [2].

In other words, CR is defined as an intelligent radio that can be programmed and configured dynamically based on interaction with its environment to achieve predefined objectives, one of which can be shared use of a spectrum band. As an intelligent radio, it has the *cognitive capability* to identify any transmission opportunity in time, frequency, and space domains through spectrum sensing, and accordingly reconfigure its transceiver parameters to communicate over the identified opportunities (*reconfigurability*) in a way to inflict minimum interference on the operation of other networks sharing the same spectrum.

Dynamic spectrum access (DSA) is one of the main paradigms of cognitive radio technology. In DSA systems, unlicensed or secondary users (SUs) are allowed to identify the under-utilized portions of licensed spectrum, often called "white holes" or "spectrum holes", and utilize them opportunistically as long as they do not cause any harmful interference to the communications of the licensed or primary users (PUs). Later on, this paradigm was extended to identification of any opportunity available within primary network operation and taking advantage of it as long as the interference impinged on the primary network is within tolerable limits.

The vast application and attention toward DSA has been so much that the terms "cognitive radio" and "DSA" are almost used interchangeably in wireless communication literature. In this paper as well, we are focusing on DSA applications of cognitive radio technology and wherever we speak of CR, we mean the spectrum sharing and DSA application of a cognitive radio system.

In order to adapt to the dynamic spectrum environment, the CRs perform following four functions [3]

*Spectrum sensing* : The first function of a CR user is to determine which portions of spectrum are available (i.e. not used by a licensed user). Therefore, it should monitor the whole available spectrum bands and identify those which are appropriate for communication.

*Spectrum allocation* : Having found the candidate bands through spectrum sensing, then a CR should select the best one and operate over it. Different policies have been proposed for this allocation process. For a successful spectrum decision and message transmission, a CR user should consider the application requirements (based on the QoS assumptions), routing metrics, and network conditions as well.

*Spectrum sharing* : As the spectrum is shared between primary and secondary users, access to spectrum should be coordinated in a way to prevent collisions between multiple CR users, or interference to primary users.

*Spectrum mobility* : As CR users are unlicensed, they should give priority to primary users and move to another spectrum hole whenever a PU requires the current spectrum band for its communication.

These four CR-specific functions are additional to the main functions (such as coding/decoding, modulation/demodulation, routing, etc.) required by any wired or wireless communication system. In a typical communication system, usually each function is implemented in a specific network layer. In a CR system, however, any of the four spectrum management functions incorporate some other functions and interactions inherent to more than a single layer. Each function influences different layers of the CR and then the operation of the whole CR network.

## II. CROSS LAYER DESIGN

Traditionally, system architectures follow strict layering principles, which ensure interoperability, flexibility, and efficient implementations. ISO reference model [4] was developed in 1980s to support standardization of network architectures using the layered model. Layered architecture follows abstraction principle, that is implementation details and internal parameters of protocols inside a layer are hidden to the remainder layers. Inter-layer communication, which are limited to procedure calls and responses, happens only between adjacent layers. Each layer makes use of the services provided by the layers below it, and in turn, makes its services available to the layers above.

In addition to modularity in protocol design which enables interoperability and flexibility, there are some other key benefits in using layering paradigms. The independence between abstraction layers allows each layer to evolve independently regardless of new or old technologies deployed in other layers. Moreover, the layered architecture models facilitate standardization.

Although standardization of layered protocol stacks has enabled fast development of interoperable systems, at the same time has limited the performance of the overall architecture, due to lack of coordination among non-adjacent layers. This issue is particularly relevant for wireless networks, where the very physical nature of the transmission medium (time-varying behavior, multipath fading, limited bandwidth, severe interference and propagation environments) introduces several performance limitations that cannot be handled well in the framework of the layered architectures.

For example, some protocols such as TCP/IP were originally designed for wired links (characterized by high bandwidth, low delay, low packet loss probability, high reliability, static routing, and no mobility), which perform poorly in wireless domain. TCP interprets the packet errors on a wireless link as network congestion (which is usually not the case in fading wireless systems) and invokes congestion control mechanisms (rather than antifading techniques) resulting in degradation of performance. This problem is often resolved by direct communication between the link layer and the transport layer.

The problem of power control, especially in ad hoc or cooperative wireless networks where there are more than two communicating nodes, is another example of layered architecture limitation. Power control clearly influences the network topology, which is a concern of the network layer. It also impacts how much spatial reuse can be achieved, that is, how far apart can two ongoing communication sessions be without interfering with each other, which is a concern of the medium access control (MAC) layer. Power control is also linked to the processing at the physical layer, because the signal processing at the physical layer determines how stringent the requirements on the power control need to be. All these factors determine the end-to-end throughput. Furthermore, the transmitted power(s) determines the lifetime of the nodes (and the network) which one would want to maximize. Hence, the problem of power control cannot possibly be handled at any one layer in isolation, as is done while designing protocols in the framework of the layered architectures.

Transmission of real time multimedia content with QoS requirements over wireless links, energy efficiency of a communication device and network security, and vertical handover in wireless networks are some other examples of problems which may not be solved with a layered architecture reference model.

To overcome such limitations, cross layer (CL) design has been proposed. It can be defined as a "protocol design by the violation of a reference layered architecture" [5]. Its core idea is to maintain the functionalities associated to the original layers but to allow coordination, interaction and joint optimization of protocols crossing different layers.

Examples of violation of a layered architecture include creating new interfaces between layers, redefining the layer boundaries, designing protocol at a layer based on the details of how another layer is designed, joint tuning of parameters across layers, and so on.

There are three main general architectures for cross layer design:

1. *Merging layers*: adjacent layers are merged into a single layer with optimum functionality of initial layers.
2. *New Interfaces*: in order to let information exchange between non-adjacent layers, new interfaces are created.
3. *Parallel or Vertical Calibration*: parameters in all or some layers are calibrated using a parallel structure as a shared interface or database to the system, which provides cross-interaction between the layers.

## CROSS LAYER DESIGN IN CR NETWORKS

While abstraction layered architecture looks like an inefficient solution in wireless links, the pursuit of cognitive networks makes it look obviously inadequate. In order to fulfill its four main functions, cognitive radios require enhanced sharing of information between all layers of the protocol stack. Spectrum sensing which can be conceptualized as the task of gaining awareness about the spectrum usage and existence of PUs in the surroundings of a CR-enabled transceiver takes place in two separate lower layers; physical (PHY) and data link layer (DLL) and requires both layers to interact with each other. Allocation of appropriate spectrum depends on the criteria spanning across all layers, such as sensing information, routing and transport requirements, cognitive application requirements (e.g. QoS, bit error rate), power restrictions and acceptable interference levels. As in spectrum sensing, spectrum sharing involves two lower layers of the ISO model; PHY and DLL. And when a CR user needs to move to another spectrum to continue its transmission (mobility), spectrum allocation procedures should be followed to pick a new channel. Therefore, as in spectrum allocation, all network layers are involved in spectrum mobility. Figure 1 illustrates four spectrum management functions spanning across different network layers.

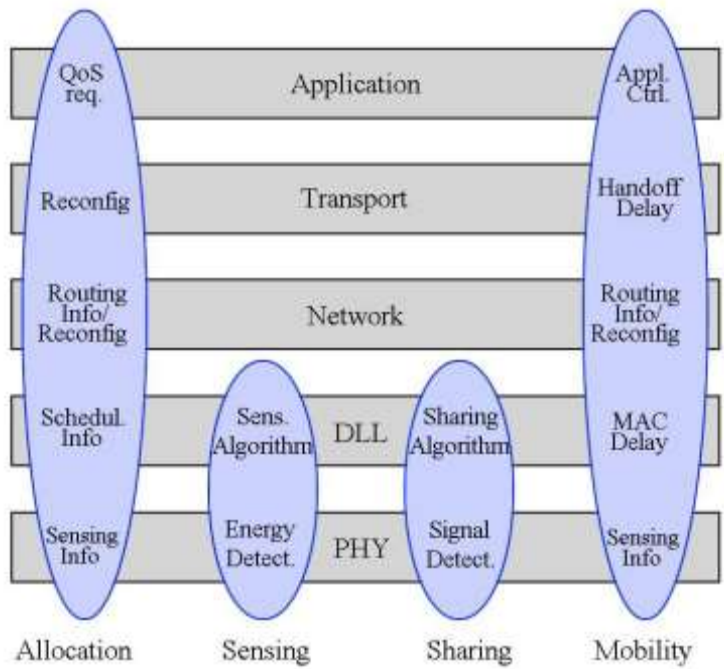

Fig. 1. Cognitive radio spectrum management framework

As we can see from Fig. 1, cognitive network not only needs to provide network functionalities to the applications, it also needs to adapt to the application needs, and to the prevailing network conditions. Such adaptability requires rich coordination between the traditional protocol layers, which is not feasible in the constraints of the layered architectures. Thus, when it comes to cognitive networks, cross-layer design is not just a matter of efficiency, but in fact it is central to the very concept of the network. There is a rich research literature of cross layer designs for cognitive radio networks, which can be found in [6-11].

## III. CROSS LAYER DESIN IN COGNITIVE RADIO STANDARDS

To be successful, any cross-layer design idea needs to be implemented and deployed in real systems. The co-design of physical layer and MAC layer has been taken into account in some of the latest wireless standards. 3G standards such as CDMA2000 have been designed with cross-layer issues in mind. In ETSI BRAN HiperLAN2 standard, the DLL and physical layer have been designed together for high throughput, low latency and QoS support. Interestingly cross-layer design ideas have been making their way into several ongoing CR standardization activities and commercial products as well, some of which are briefly outlined in this section.

### *IEEE 802.22 WRAN*

IEEE 802.22 was the first centralized cognitive radio PHY-MAC layer standard for wireless regional area networks (WRANs) developed to exploit vacant TV spectrum bands, or TV white spaces (TVWS). Its main idea was to provide a wireless broadband access in rural areas. The network in IEEE 802.22 operates in a point to multipoint basis and the CR network is divided into cells, each one comprised of a base station (BS), covering an area of radius spanning from 17 km to 100 km, and several WRAN end users, here denoted as customer premise equipment (CPEs). In addition, the standard provides PU protection through spectrum sensing and geolocation database for PU-SU coexistence, and also supports self-coexistence with other WRANs through the coexistence beacon protocol (CBP).

IEEE 802.22 incorporates advanced cognitive radio capabilities including dynamic spectrum access, incumbent database access, accurate geolocation techniques, spectrum sensing, regulatory domain dependent policies, spectrum etiquette, and coexistence for optimal use of the available spectrum. The reference architecture for IEEE 802.22 systems addresses PHY and MAC layers and the interfaces to a station management entity (SME) through PHY and MAC layer management entities (MLMEs), as well as to higher layers such as Internet protocol (IP) layer.

Similar to IEEE 802.11, SME in 802.22 is a cross layer entity which communicates with multiple protocol layers. SME is in charge of managing the internal processes carried out within different layers of protocol stack. The functional blocks are interconnected through service access points (SAPs) that allow the exchange of information between layers and entities in a cross layer manner.

### *IEEE 802.11af, White-Fi*

IEEE presented in 2014 another standard to exploit the TVWSs in a more domestic scenario (personal and portable devices). The IEEE 802.11af, or White-Fi, adapts the current IEEE 802.11 standards to make use of the TVWSs between 54 and 750 MHz, just as in the IEEE 802.22, but by considering a smaller transmission range (up to 1 km). These modifications occur mainly in PHY and MAC layers of legacy IEEE 802.11 standards.

In contrast to the legacy IEEE 802.11 standards with two types of main components (access point and stations), IEEE 802.11af comprises of the following four main components:

- Geolocation database (GDB): A database which stores the permissible frequencies and operating parameters mandated by regulatory authorities for different geographic locations. Any white-fi network should contact a GDB prior to and during its operation to be informed of the available TVWSs and permissible operating parameters.

- Registered location secure server (RLSS): This server acts as a local database that contains the geographic location and operating parameters for a small number of basic service sets (BSSs). The operation of the access points and wireless users in a given BSS is stored in the RLSS.

- Geolocation database dependent enabling station (GDD-Enabling STA): This component corresponds to the AP of legacy IEEE 802.11 standards and controls the operation of other stations within its service area. The GDD-enabling STA communicates with GDB to get the available TVWS information, and updates RLSS with the acquired info about location of active stations within its BSS service area.

- Geolocation database dependent station (GDD-Dependent STA): The GDD dependent STAs are equivalent to the STAs in the BSS architecture of legacy IEEE 802.11 standards. GDD-dependent STAs are allowed to use the available TVWS channels under the control of a GDD-enabling STA. The corresponding information about the permissible operating frequencies and parameters are provided by GDD enabling STA to GDD dependent STAs through a White Space Map (WSM).

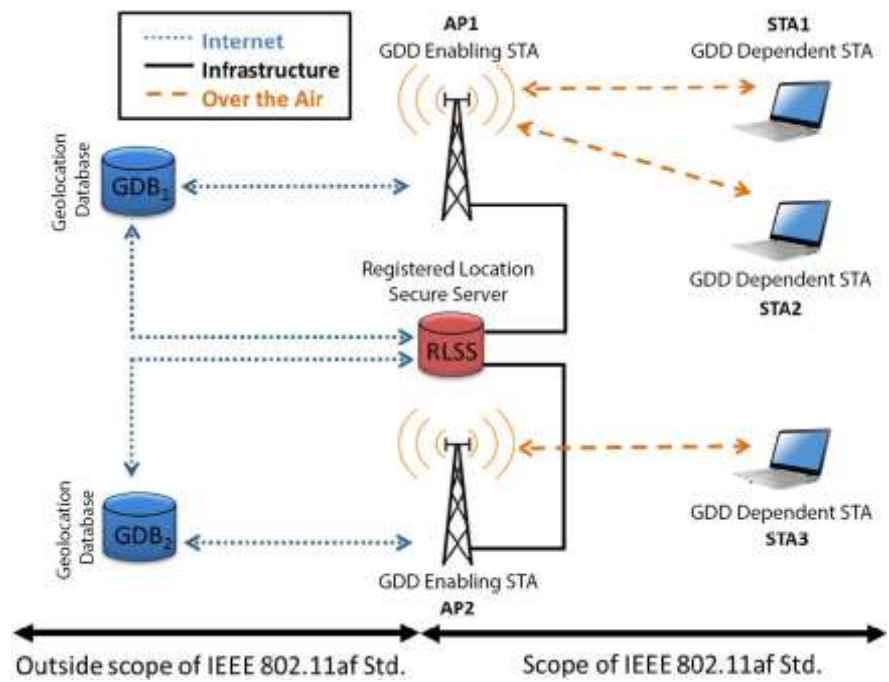

Fig.2. Example TVWS network including all 802.11af architecture entities [12].

The IEEE 802.11af standard only defines the communication protocol between the GDD-dependent STAs, GDD-enabling STAs, and RLSS. Communication between GDB and GDD-enabling STAs or RLSS is beyond the scope of IEEE 802.11af. Architecture of IEEE 802.11af is shown in Fig.2.

As IEEE 802.11af is an adaptation of legacy 802.11 standards, it inherits the cross-layer design implementations therein. For example, station management entity (SME) in 802.11 is cross-layer entity which exists in IEEE 802.11af as well. The cross-layer solutions deployed in 802.11e standard to ensure QoS which may be considered in later revisions of 802.11af for QoS support. In addition, the stations in 802.11af standard are cognitive devices which employ CL methods to sense and allocate appropriate spectrum. With cross-layer communication between the network layer and the spectrum databases, information from the network layer can influence the spectrum allocations. The network layer can inform the database about traffic volumes and possibly the location of the important or heavy users of network services. The database can then allocate more frequencies to the parts of the network with large amount of traffic.

*IEEE 802.21*

The problem of vertical handovers between heterogeneous technologies is being addressed by the IEEE 802.21 standard. Vertical handovers refer to the automatic switching from one technology (e.g. IEEE 802) to another (e.g. cellular technology) without service interruption. Vertical handover decisions (handover metrics), in contrary to those in horizontal handovers, which are mainly based on received signal strength (RSS), should include RSS, user preference, network conditions, application types, cost etc. which are considered in different layers. Mobility management (functionalities required to provide session continuity) is another important problem in vertical handovers which may be solved in different layers such as application layer, network layer, transport layer, etc. in a cross layer manner. In the cross layer-based approaches, fewer handover delays, less packet loss, higher throughput, and better quality of service are attained.

IEEE 802.21 standard is an interesting example in which an explicit cross-layer entity is introduced and deployed. It provides a framework for cross-layer signaling that allows higher levels to interact with lower layers to provide session continuity without dealing with the specifics of each technology. In IEEE 802.21 a mid-layer between layers 2 and 3 is inserted (layer 2.5) through which upper layer services known as media independent handover (MIH) users communicate with protocols of the lower Link layer and Physical layer.

The current published version of IEEE 802.21 standard is not drafted as a cognitive radio standard. However, in order to be a comprehensive standard for seamless communication over different wireless technologies, it goes without saying that inclusion of some cognitive functionalities for supporting CR networks (e.g. IEEE 802.11af or IEEE 802.22) would be considered in its future versions.

*IEEE 802.16h-CognitiveWiMAX*

IEEE 802.16, also denoted as Worldwide Interoperability for Microwave Access (WiMAX), in its second amendment (denoted by IEEE 802.16h in 2009, and superseded by IEEE 802.16-2012 edition in 2012) introduced some cognitive mechanisms for license-exempt operation of WiMAX networks for frequencies below 11GHz. In a real scenario a WiMAX network may coexist with licensed users (denoted as specific spectrum users) and other unlicensed users (denoted as non specific spectrum users) sharing the same frequency band. In such case, the interference that the WiMAX system may cause to each of the users can be different.

IEEE 802.16h proposes a cognitive PHY-MAC framework to achieve two major goals; first to enable the coexistence among license exempt (LE) systems based on IEEE 802.16 standard and second, to facilitate the coexistence of such systems with primary users in a licensed system of any type. It defines two modes of operation: 1) uncoordinated coexistence mechanisms (WirelessMAN-UCP) which does not require much interaction among the different licensed or unlicensed systems, and 2) coordinated coexistence mechanisms (WirelessMAN-CX), in which multiple CR networks coexist in the same region. in order to reduce the generated interference, these networks are related to some neighbouring groups within which they coordinate their transmissions based on acceptable interference limits.

*ECMA -392:*

ECMA International is an industry association founded in 1961 and dedicated to the standardization of information and communication technology (ICT) and consumer electronics (CE). ECMA-392 standard titled "MAC and PHY for Operation in TV White Space" is a cross layer standard published in Dec 2009. Its target applications are wireless home and business network access over TV white spaces, similar to IEEE 802.11af.

The major difference between two standards are the mechanisms of incumbent protection and supported bandwidth. In addition to acquiring list of available channels through a database, ECMA-392 additionally supports spectrum sensing functionality and periodically checks the presence or an incumbent signal over the channel in use, which is not supported in IEEE 802.11af yet.

ECMA-392 standard specifies a MAC sub-layer and a PHY layer for personal/portable cognitive wireless networks operating in TV bands. This standard also specifies a MUX (a session management protocol) sub-layer which is a cross layer entity enabling the coexistence of concurrently active higher layer protocols within a single device. This sublayer routes outgoing and incoming MAC service data units between their corresponding higher layers and the two lower layers. It also specifies a number of incumbent protection mechanisms which may be used to meet regulatory requirements.

## IV. CROSS LAYER DESIGN IDEAS BASED ON CR STANDARDS

Standards usually set requirements, specifications, guidelines and characteristics to be observed consistently by designers and manufacturers, and usually do not specify algorithms, methods or protocols for this purpose. Designers have the freedom to use or propose any algorithm or method as long as they meet the specified requirements of the respective standard. For example, IEEE 802.22 standard considers two types of sensing (fast sensing and fine sensing) and the specifications of each type such as corresponding sensing duration and level of accuracy are mentioned in it, but no sensing algorithm is specified, although details of a number of spectrum sensing algorithms are included in an informative annex.

When a standard is released, designers try to propose efficient methods and algorithms within the framework of the standard to fulfill the functions specified in the standard efficiently, among which CL design ideas have been of great attention for cognitive radio standards.

In this section we briefly explain some of the CL works based on published CR standards:

As mentioned in section II, TCP performance in wireless networks has been an issue which motivates designers to seek a cross layer solution for it. Authors in [13] presented an analysis of TCP performance in IEEE 802.22 WRAN based on cognitive radio networks and have shown that TCP performance is affected adversely due to PU activity in a WRAN, DSA by the un-licensed users and associated quiet periods that can be comparable with RTO interval of the sender's TCP, a scenario which can trigger TCP's congestion control mechanism un-necessarily and fruitlessly. They proposed two cross-layer solutions having close interaction between the transport and MAC/PHY layers in order to improve the communication efficiency in a CRN. The first approach makes base station resort to local recovery of lost frames between CRN base station and its clients, while the second approach implements a modified split TCP connection in which the base station sends crafted acknowledgements back to an Internet-side host on behalf of the corresponding CRN client to boost transmission speed.

Sensing the incumbent signal is an important task in all cognitive radio paradigms, as in WRAN. IEEE 802.22 standard has set the requirements for detection accuracy and acceptable interference, and designers may decide their best solutions to meet these criteria. In [14] a cross-layer design (CLD) of carrier sense multiple access with collision avoidance (CSMA/CA) mechanism at the MAC layer with spectrum sensing (SpSe) at the physical layer is proposed for identifying the occupancy status of TV bands. The proposed CLD relies on a Markov chain model with a state pair containing both the SpSe and the CSMA/CA from which the collision probability and the achievable throughput were derived. The proposed method is compatible with IEEE 802.22 and ECMA-392 standards.

Another solution to improve PU protection and sensing performance in CR over TVWS is proposed in [15]. Their work investigates a cross layer mechanism called the cross layer cognitive engine (CLCE) within IEEE 802.22 standard framework which shares information between the MAC and PHY layers, so sensing measurements can influence spectrum access decisions. The CLCE forms the basis of a new enhanced detection algorithm (EDA) which defines the way a TVWS channel is accessed. The EDA utilizes the patterns in which the digital TV frequencies are deployed to determine whether a PU is occupying a channel, and importantly utilizes an energy detector which exploits local real-time measurements in the decision making. It is claimed in [15] that EDA consistently outperforms existing PU detection algorithms when applying the IEEE 802.22 WRAN standard of 90% for detection and 10% false detection thresholds.

Cognitive radio ad hoc (CRAHN) are of great interest as they are applied in sensor networks, machine-to-machine (M2M) communications and internet of things (IoT). In multihop CRAHN there are many issues which affect the network performance. Collisions due to channel contention among different links at the MAC layer, congestion due to sharing of common links among greedy sources at the transport layer, and interference due to co-utilized dynamic spectrum sharing among several simultaneous transmissions at the physical layer are key obstacles that can fundamentally reduce the network performance. In [16] two OSA-based cross-layer MAC schemes for multi-hop CRAHNs are proposed, in which the interactions of three barriers (i.e., congestion, collision, and interference) are carefully considered to seek an optimal operating point for cognitive users. The proposed cross-layer MAC protocols aim to achieve energy efficiency and contention fairness through a unified cross-layer optimization framework according to guidelines and specifications of IEEE 802.22 standard.

Supporting QoS requirements of multimedia applications over CR network is a difficult task which requires a sustained throughput to maintain acceptable video quality. This is not easily achieved in a CR network where availability to spectrum is not always guaranteed. This can be quite difficult to guarantee in a DSA environment characterized by spatiotemporal variations in spectrum availability. [17] provides a framework for opportunistically transporting heterogeneous traffic, which includes high-bandwidth media streams as well as best-effort flows, over TVWS without interfering with the operation of PUs. They designed a QoS-aware parallel sensing/probing architecture called QASPA and a cross-layer protocol which exploits inherent channel and user diversities exhibited by the wireless system. The proposed cross layer protocol uses an adaptive framing structure to minimize the control overhead, and a novel spectrum assignment strategy targeted at improving the spatial reuse of the network. Their scheme shows the superior performance of QASPA compared to the scheme used in the ECMA-392 standard.

Another effort to support transmission of multimedia content over CR network is given in [18]. The authors proposed cross layer techniques for transmission of multimedia content to a mobile station using a combination of mobile/cellular networks operating over licensed and TVWS spectrum. In their patented method scalable video coding has been used to transmit a baseline layer of multi-media content using the mobile/cellular network and one or more enhancement layers over the TVWS channels (cognitive radio). The proposed patent is compliant with 802.16, 802.11af, and 802.22 standards.

## V. CONCLUSION

Cross-layer design in cognitive radio systems has attracted much attention from academia and industry, and similar to non-cognitive radio systems it has been proved to achieve significant gain in performance over the layered design in isolation. There has been extensive study on joint design methods on lowers layers (PHY, MAC and Network) for optimizing throughput or other performance factors of a CR network. Some application driven designs involve consideration of higher layers for QoS provisioning in CR networks. Such modifications and optimizations are accomplished through introduction of new methods of spectrum sensing, sharing or allocation. Successful cross-layer design ideas have made their way into several wireless communication standards, among which CR standards are remarkable examples. As any standard is being edited and improved alongside introduction of new technologies and methods, there are many interesting research areas for efficient cross layer design ideas based on available and ongoing CR standards.